# Exceptional points make an astroid in non-Hermitian Lieb lattice: evolution and topological protection

Yi-Xin Xiao[1,2], Kun Ding[2,3], Ruo-Yang Zhang[2], Zhi Hong Hang[1,4], C. T. Chan[2*]

[1]School of Physical Science and Technology, Soochow University, Suzhou 215006, China

[2]Department of Physics, Hong Kong University of Science and Technology, Hong Kong, China

[3]The Blackett Laboratory, Department of Physics, Imperial College London, London SW7 2AZ, United Kingdom

[4]Institute for Advanced Study, Soochow University, Suzhou 215006, China

*Corresponding author: phchan@ust.hk

An astroid-shaped loop of exceptional points (EPs), comprising four cusps, is found to spawn from the triple degeneracy point in the Brillouin zone (BZ) of a Lieb lattice with nearest-neighbor hoppings when non-Hermiticity is introduced. The occurrence of the EP loop is due to the realness of the discriminant which is guaranteed by the non-Hermitian chiral symmetry. The EPs at the four cusps involve the coalescence of three eigenstates, which is the combined result of the non-Hermitian chiral symmetry and mirror-T symmetry. The EP loop is exactly an astroid in the limit of an infinitesimal non-Hermiticity. The EP loop expands from the $M$ point with increasing non-Hermiticity and splits into two EP loops at a critical non-Hermiticity. The further increase of non-Hermiticity contracts the two EP loops towards and finally to two EPs at the $X$ and $Y$ points in the BZ, accompanied by the emergence of Dirac-like cones. The two EPs vanish at a larger non-Hermiticity. The EP loop disappears and several discrete EPs are found to survive when next-nearest hoppings are introduced to break the non-Hermitian chiral symmetry. A topological invariant called the discriminant number is used to characterize their robustness against perturbations. Both discrete EPs and those on the EP loop(s) are found to show anisotropic asymptotic behaviors. Finally, the experimental realization of the Lieb lattice using a coupled waveguide array is discussed.

# I. introduction

The occurrence of exceptional points (EPs) is the most prominent phenomenon in non-Hermitian systems among others [1–9], which have been investigated in optics [10,11], acoustics [12,13], electric circuits [14,15], to name a few. EPs have brought about a wealth of promising applications, such as lasing [16–18] and enhanced sensitivity [9]. Many recent works attempted to extend the topological band theory to non-Hermitian Hamiltonians [19–25]. Apart from the ubiquitous doubly degenerate EPs, higher-order EPs [9,26,27] have also been studied extensively.

EPs may form a continuum of various shapes, such as EP lines [10,28], rings [11,29–32], ellipses [26], surfaces [33–35] and nexuses [36]. Exceptional rings were found to emerge from non-defective degeneracy points (NDPs) such as two-fold degenerate Weyl points when non-Hermiticity is introduced [11,29,37,38]. It is intriguing to understand how a higher order NDP is affected by non-Hermiticity, especially whether a higher-order EP ring will form. This motivates us to explore the non-Hermitian Lieb lattice [39–41] which possesses a triple degeneracy in the Hermitian scenario.

In the Lieb lattice with only nearest-neighbor (NN) hoppings included, we found that the exceptional points spawn from the triple degeneracy point (DP) and form a closed loop, which takes an astroid shape when the non-Hermiticity is small. In particular, four EPs of order 3 are found at the cusps of the astroid. The EP loop and EPs of order 3 are protected by the non-Hermitian chiral symmetry. Increasing non-Hermiticity leads to the expansion and splitting of the EP loop into two loops. Further increase in non-Hermiticity induces the two EP loops to contract towards and finally vanish at the Brillouin zone (BZ) boundary. The introduction of next-nearest-neighbor (NNN) hoppings turns the EP loop to several discrete EPs, which show robustness against various perturbations. The robustness is guaranteed by a nonzero discriminant number, which is a topological invariant defined for an EP [27].

The work is organized as follows. Sec. II covers results for the system with NN hoppings, and Sec. III takes NNN hoppings into account. In Sec. II, we will see that the Lieb lattice exhibits many interesting results, such as EP loop with cusps, interesting evolution of EP trajectories, Dirac-like cones, and anisotropic EPs. In Sec. III, we will show the existence of discrete EPs in the presence of NNN hoppings and their interesting evolution behaviors. The discriminant number, a generalized winding number, is used to explain the robustness and describe the relevant creation/annihilation behaviors of EPs. In Sec. IV, we will discuss the effect of perturbations on both the EP loop and discrete EPs. In Sec. V, we discuss the experimental relevance of the model systems. Finally, we conclude our results in Sec. VI.

## II. Non-Hermitian Lieb lattice with NN hoppings
### A. Exceptional loop with cusps

The Lieb lattice with balanced gain and loss ($\pm i\gamma$) is shown in Fig. 1(a). Three sites in the unit cell are labeled as $A$, $B$, $C$. The NN and NNN hoppings are denoted by $t$ and $w$. We assume a unity lattice constant, $a = 1$, and that the lattice is periodic in the $xy$-plane. The Bloch Hamiltonian is

$$H(\vec{k}) = \begin{pmatrix} i\gamma & T_{AB} & T_{AC} \\ T_{AB} & 0 & T_{BC} \\ T_{AC} & T_{BC} & -i\gamma \end{pmatrix}, \tag{1}$$

where $T_{AB} = 2t\cos\left(\frac{k_y}{2}\right)$ and $T_{BC} = 2t\cos\left(\frac{k_x}{2}\right)$ and $T_{AC} = 4w\cos\left(\frac{k_x}{2}\right)\cos\left(\frac{k_y}{2}\right)$. Without loss of generality, $t = 1$ is assumed. In this section only NN hoppings are considered, namely $w = T_{AC} = 0$. The $w \neq 0$ scenario will be discussed in Sec. III. For convenience we will call the Lieb lattice with (without) NNN hoppings as the NNN (NN) system.

The band structure for the Hermitian system with $\gamma = 0$ is shown in Fig. 1(b), where the BZ is chosen to be $\vec{k} \in [0,2\pi] \times [0,2\pi]$ and a triple degeneracy occurs at $M = (\pi, \pi)$ due to the chiral symmetry and $C_{4v}$ symmetry. The symmetry in the band structure with respect to the $E = 0$ flat band is due to the chiral symmetry.

The band structure for $\gamma = 2$ is shown in Figs. 1(c) and 1(d). The colors in Fig. 1(d) are taken in accordance with Fig. 1(c), and they exhibit discontinuities due to the presence of branch cuts. For clarity, only a quarter of BZ is shown in Fig. 1(d) using the fact that $E(-k_x, k_y) = E(k_x, k_y) = E(k_x, -k_y)$ due to mirror symmetries [42]. An EP loop with four cusps [43,44] is centered at $M$ and consists of four arcs, two cyan and two magenta. This is in stark contrast to our expectation that an EP ring may occur [11]. The four cusps marked by gray dots are EPs of order 3, and all other points on the arcs are EPs of order 2. We refer to them as EP3s and EP2s and two EP2 arcs coalesce at a cusp to form an EP3. The degenerate eigenvalue corresponding to an EP2 on the cyan (magenta) arcs satisfies $Im(E) > 0$ ($Im(E) < 0$) as is seen in Fig. 1(d).

Detailed analysis shows that an infinitesimally small $\gamma$ would turn the triple degeneracy point into a vanishingly small loop of EPs [42], and the shape of the loop is found to be a special planar curve called an astroid [42]. The EP loop expands from $M$ when $\gamma$ is increased gradually as is shown in Fig. 2(a) where several $\gamma$ values are labeled. To supplement Fig. 1 (d), the imaginary part of eigen-energies along the EP loop for $\gamma = 2$ is shown in Fig. 2(b), where $k_1, k_2, k_3, k_4$ denote the four EP3 cusps labeled in Fig. 2(a). The three bands are plotted by green solid, blue dashed, red solid curves, respectively. The degeneracies signify the existence of EPs.

Compared to an EP2, more constraints need to be satisfied so that an EP3 exists. The explicit conditions for their existence will be provided shortly.

### B. Origin of EP loop and EP3s

The NN system has non-Hermitian chiral symmetry $\Sigma^{-1}H^\dagger(\vec{k})\Sigma = -H(\vec{k})$ [42], which is different from its Hermitian counterpart due to $H^\dagger(\vec{k}) \neq H(\vec{k})$ [45]. Here $\Sigma = diag(1,-1,1)$ is a diagonal matrix. The non-Hermitian chiral symmetry dictates that the eigenvalues take the form as either

$$E_{1,3} = \pm a + ib, E_2 = ic, \qquad (2)$$

or

$$E_m = i\lambda_m, m = 1, 2, 3, \qquad (3)$$

where $a, b, c, \lambda_m$ are real [42]. As a consequence, the characteristic polynomial $f(E,\vec{k}) = \det[E - H(\vec{k})]$ takes the form $f(E,\vec{k}) = \prod_{i=1}^{3}[E - E_i(\vec{k})] = E^3 + a_1 E + a_0$, where $a_1$ and $a_0$ are real and imaginary, respectively. The absence of the $E^2$ term in $f(E,\vec{k})$ is due to the zero trace of $H(\vec{k})$, namely $\sum_i E_i(\vec{k}) = 0$. For the non-Hermitian Lieb lattice we have

$$f(E,\vec{k}) = E^3 + a_1(\vec{k})E + a_0(\vec{k}), \qquad (4)$$

where

$$a_1(\vec{k}) = \gamma^2 - (T_{AB}^2 + T_{BC}^2), \quad a_0(\vec{k}) = i\gamma(T_{BC}^2 - T_{AB}^2). \qquad (5)$$

In compact notation we can write $f(E,\vec{k}) = \sum_{m=0}^{3} a_m(\vec{k})E^m$ where $a_3(\vec{k}) = 1$ and $a_2(\vec{k}) = 0$.

The locations of EPs in the BZ can be directly determined by solving $\Delta(\vec{k}) = 0$, where $\Delta(\vec{k}) = \prod_{i<j}[E_i(\vec{k}) - E_j(\vec{k})]^2$ is the discriminant of $f(E,\vec{k})$. This method can also locate where the non-defective degeneracies points (NDPs) occur. The points determined are EPs if two or more eigenvectors coalesce. From Eqs. (2) and (3) we know that $\Delta(\vec{k})$ is purely real, which underlies the existence of EP loop and the EP3s. Explicitly the discriminant $\Delta(\vec{k})$ is found to be [42]

$$\Delta(\vec{k}) = -4a_1^3 - 27a_0^2. \tag{6}$$

The $\vec{k}$ points that satisfy $\Delta(\vec{k}) = 0$ trace out the EP loops in Fig. 2(a). There are exceptional cases when the solutions to $\Delta(\vec{k}) = 0$ correspond to non-defective degeneracies [42]. The order (multiplicity) of the EPs (NDPs), 3 or 2 here, remains to be determined by additional constraints. We observe that the only possible triple root of $f(E, \vec{k})$ in Eq. (4) is $E = 0$. Consequently we require that $a_1(\vec{k}) = 0$ and $a_0(\vec{k}) = 0$. Obviously $\Delta(\vec{k}) = 0$ and $a_1(\vec{k}) = 0$ implies $a_0(\vec{k}) = 0$. Therefore, the constraints for the existence of a triple DP (including an EP3) are simply

$$\Delta(\vec{k}) = 0 \text{ and } a_1(\vec{k}) = 0, \tag{7}$$

where $\Delta(\vec{k})$ has the explicit form,

$$\Delta(\vec{k}) = 4(T_{AB}^2 + T_{BC}^2 - \gamma^2)^3 + 27\gamma^2(T_{BC}^2 - T_{AB}^2)^2. \tag{8}$$

A point $\vec{k}$ determined by $\Delta(\vec{k}) = 0$ and $a_1(\vec{k}) \neq 0$ corresponds to a double degeneracy. A double (triple) DP is an EP2 (EP3) if the 2 (3) eigenvectors associated with the DP coalesce to one eigenvector. In this way we determined the EP loops and EP3s shown in Fig. 2(a).

We note that $\Delta(\vec{k})$ is generally complex. The realness of $\Delta(\vec{k})$ in Eq. (6) is crucial for the existence of EP2 continua and EP3s in the non-Hermitian Lieb lattice. EP2s can exist and form continua for appropriate system parameters because the constraint (i.e., $\Delta = 0$) involved is outnumbered by the degrees of freedom (i.e., $k_x, k_y$). In contrast, the existence of EP3s necessitates one more constraint, namely $a_1(\vec{k}) = 0$. As a consequence, EP3s only exist at four discrete $\vec{k}$ points $(\pi \pm q, \pi \pm q)$, where $q = 2\sin^{-1}\left(\frac{\sqrt{2}\gamma}{4}\right)$. The function $\sin^{-1}\left(\frac{\sqrt{2}\gamma}{4}\right)$ indicates that EP3s cannot exist when $\gamma > 2\sqrt{2}$. All the EP3s are located at the two BZ diagonals $k_y = k_x$ and $k_y = 2\pi - k_x$, which is protected by the non-Hermitian chiral symmetry and mirror-T symmetries [42].

EP3s occur as special solutions to the EP2 constraint $\Delta(\vec{k}) = 0$; therefore, EP3s join the EP2 arcs. The contour lines for $\Delta(\vec{k}) = 0$ and $a_1(\vec{k}) = 0$ are illustrated in the Supplemental Material [42].

Since $\Delta(\vec{k})$ is a periodic function of $\vec{k}$ for the NN system, the EPs given by the contour line $\Delta(\vec{k}) = 0$ must form one or several loops in the BZ if solutions exist, as is demonstrated in Fig. 2(a). There are also situations when EP2s or EP3s do not exist if the relevant constraints could not be satisfied by any $\vec{k}$ in the BZ.

In short, the existence of the EP loop and EP3s is due to the realness of the discriminant which is guaranteed by the non-Hermitian chiral symmetry. We noticed a recent work which shows exceptional rings protected by the non-Hermitian chiral symmetry in correlated systems [30].

We note that the non-Hermitian chiral symmetry, i.e., $\Sigma^{-1}H^\dagger(\vec{k})\Sigma = -H(\vec{k})$, is preserved in the more general situation when the three onsite energies $\epsilon_A, \epsilon_B, \epsilon_C$ take arbitrary imaginary values [42]. The characteristic polynomial generally contains the $E^2$ term since the trace of $H(\vec{k})$ is nonzero. EP3s exist when the constraints $\Delta(\vec{k}) = 0$ and $a_2^2(\vec{k}) - 3a_1a_3(\vec{k}) = 0$ can be simultaneously satisfied in the BZ [42]. The presence of mirror-T symmetries is not necessary for the existence of EP3s. The NN system with $\epsilon_A = -\epsilon_C = i\gamma, \epsilon_B = 0$ is thus a subset of chiral-symmetric systems that may admit EP loops and EP3s [42].

### C. Evolution of EP loop as non-Hermiticity increases

Previously we mentioned the expansion of the EP loop with increasing $\gamma$. A transition occurs at $\gamma = 2\sqrt{2}$ when the four EP3s converge to a single EP3 at $\Gamma$. Further increase in $\gamma$ splits the EP loop centered at $M$ into two small loops centered at $X = (\pi, 0)$ and $Y = (0, \pi)$, respectively. The EP3s vanish for $\gamma > 2\sqrt{2}$ because Eq. (7) is no longer satisfied by any $\vec{k}$ in the BZ. From continuity we know that the two split loops also have cusps where they intersect $k_y = 0$ and $k_x = 0$. Then with $\gamma$ increasing, the two loops contract until they shrink to two discrete EP2s at $X$ and $Y$ when $\gamma = 4$. There are no longer EPs for $\gamma > 4$ because $\Delta(\vec{k}) = 0$ cannot be satisfied in the BZ. Thus $\gamma = 2\sqrt{2}$ and $\gamma = 4$ are two transition points, where the number of EP loops changes from 1 to 2 and then to 0. The EP loop and its evolution exhibit interesting phenomena such as anisotropic

EPs and Dirac-like cones which will be shown shortly. We note that the tuning of $\gamma$ could be viewed as one type of perturbations that respect the non-Hermitian chiral symmetry, which only change slightly but do not affect the existence of the EP loop and EP3s. In this sense, the EP loop and EP3s have topologically protected robustness against perturbations that preserve the non-Hermitian chiral symmetry [46].

### D. Anisotropic EP2s and EP3s

The EPs within a continuous EP curve could exhibit anisotropic behaviors [13,26,47]. That is, the asymptotic behavior of the associated eigenvalues near an EP along the tangent line to the EP curve is different from any other direction (including the normal direction). They are also called hybrid points [21,48,49]. We know that the tangent path traversing the EP is special in that no transition between the exact phase and broken phase would occur at the EP. The two sides of the EP correspond to the same phase, exact or broken.

All points on the EP2 curves are anisotropic EP2s. For illustration, we focus on the EP2 at $\vec{k} = (4/3, 1)\pi$, which is the intersection of the navy line and the orange line marked in Fig. 3(a). The navy line and the orange line represent the normal and the tangent paths, respectively. The two paths are labeled by "b" and "c" to indicate their association with the band structures shown in Figs. 3(b) and 3(c), where $\delta k$ denote the deviations from the EP2 along the two paths. The familiar square-root bifurcation in a pair of eigenvalues is seen in Fig. 3(b). In contrast, the same pair of eigenvalues exhibits linear crossing in Fig. 3(c). We found analytically that $E_{1,2} = -i \pm 3^{1/4}\sqrt{\delta k} + O(\delta k^{3/2})$ with $\delta k = k_x - 4\pi/3$ near the EP2 in Fig. 3(b), and $E_{1,2} = -i \pm i\delta k/\sqrt{3} + O(\delta k^2)$ with $\delta k = k_y - \pi$ near the EP2 in Fig. 3(c). We also confirmed that an arbitrarily chosen EP2 on the loop also show anisotropic behaviors in eigenvalues.

The EP3s also exhibit anisotropy. We take the EP3 at $\left(\frac{1}{2}, \frac{1}{2}\right)\pi$ for illustration. The tangent path ($k_y = k_x$) and the normal path ($k_y = \pi - k_x$) at the chosen EP3 are represented by the red line and green line, respectively. The eigenvalues near the EP3 are asymptotically $E \propto \delta k^{1/3}$ and $E \propto \delta k^{1/2}$, respectively, along the green path and red path, as is shown in Figs. 3(d) and 3(e), where $\delta k$ denotes the deviation from the EP3 along each corresponding path. In contrast to the anisotropic EP2s discussed previously, it is the tangent path at the EP3 that makes the transition from the exact phase to broken phase. This distinguishes the singular EP cusps from other smooth EP curves such as EP rings [11] and ellipses [26], for which both sides of an EP along the tangent path correspond to the same phase. The anisotropic behavior of EP3 here is reminiscent of the arbitrary-order EPs in Ref. [26], where eigenvalues asymptotically take square-root and linear forms, respectively.

Phase rigidity can well characterize the coalescence (or splitting) of eigenstates near an EP [26,50]. It is defined as

$$\rho_m = |\langle \psi_m^L | \psi_m^R \rangle|, \tag{9}$$

where $\langle \psi_m^L|$ and $|\psi_m^R\rangle$ are self-normalized left and right eigenvectors for the $m$-th state. We note that $\rho = 0$ for eigenstates associated with an EP and $\rho = 1$ for eigenstates of a Hermitian system; and $0 \leq \rho \leq 1$ for a general eigenstate in non-Hermitian systems. Phase rigidity adheres to a power law $\rho_m \propto |\delta|^\chi$ near an EP, where $\delta$ denotes a small deviation from the EP and the exponent $\chi$ depicts the rate that $\rho_m$ vanishes when approaching an EP. In other words, $\chi$ characterizes the rate that two or more eigenstates turn aligned when approaching an EP.

For the four paths in Fig. 3(a), phase rigidity is calculated for an arbitrarily chosen eigenstate related to the EP2/EP3 and shown in Fig. 3(f). The phase rigidity vanishes at the EP2/EP3 for all four paths. The phase rigidities obey $\rho \propto \delta k^{1/2}$ and $\rho \propto \delta k$ for the EP2-related eigenstates along paths corresponding to Figs. 3(b) and 3(c), respectively. The results agree with the power laws $\rho \propto \delta^{(N-1)/N}$ and $\rho \propto \delta^{N-1}$ reported earlier for an EP of order $N$ (EPN) for which the eigenvalues follow asymptotically $E \propto \delta^{1/N}$ and $E \propto \delta$, respectively, where $\delta$ is the deviation [26]. For the EP3, the power laws $\rho \propto \delta k^{2/3}$ and $\rho \propto \delta k$ along the green and red paths correspond to $\rho \propto \delta^{(N-1)/N}$ and $\rho \propto \delta^{(N-1)/2}$ reported for an EPN follow $E \propto \delta^{1/N}$ and $E \propto \sqrt{\delta}$ behaviors, respectively [26].

### E. Dirac-like cones residing at discrete EPs

We also found a new phenomenon that the imaginary part of energy sheets exhibit Dirac-like cones which reside at the two discrete EPs, i.e., $X$ and $Y$, when $\gamma = 4$, as is shown in Fig. 4(a). The BZ region is chosen as $\left[-\frac{\pi}{2}, \frac{3\pi}{2}\right] \times \left[-\frac{\pi}{2}, \frac{3\pi}{2}\right]$ for a better view. The two discrete EPs come from the shrinking of the two EP loops centered at $X$ and $Y$ when $2\sqrt{2} < \gamma < 4$. Different from the Dirac cones in systems such as graphene, EPs appear at the cone vertices and there is only a single eigenstate associated with the degeneracy point. We refer to the cones as defective cones. The purely imaginary eigenvalues are due to the non-Hermitian chiral symmetry [42].

We focus on the cone at $X$. The effective Hamiltonian near $X$ takes the form

$$H_{eff}(\vec{q}) = \begin{pmatrix} 4i & 2 - \frac{q_y^2}{4} \\ 2 - \frac{q_y^2}{4} & -i\frac{q_x^2}{6} \end{pmatrix}, \tag{10}$$

where $\vec{q} \equiv (q_x, q_y) = \vec{k} - X$, with $|q_x|, |q_y| \ll 1$. The eigenvalues are

$$E_\pm(\vec{q}) = i\left(2 \pm \sqrt{\frac{1}{3}q_x^2 + q_y^2}\right), \tag{11}$$

where linear terms of $q_x, q_y$ are kept. The eigenvalues near the EP2 at $X$ follow $E \propto |\vec{q}|$ asymptotically in all directions, in contrast to the anisotropic EPs discussed in Sec. II.D. And the phase rigidity is found to follow $\rho(\vec{q}) \propto |\vec{q}|$ isotropically.

Another difference between the defective Dirac-like cones here and Dirac cones in graphene is that the Berry phase for the EP is 0 rather than $\pi$, as is calculated using $\Phi = i \oint_C \langle u^L(\vec{k}) | \nabla_{\vec{k}} u^R(\vec{k}) \rangle \cdot d\vec{k}$, where $\langle u^L(\vec{k})|$ and $|u^R(\vec{k})\rangle$ denote the left and right eigenvectors of $H(\vec{k})$ and $C$ denotes a small loop on the cone around the vertex.

For comparison, the band structure for $\gamma = 4.01$ is shown in Fig. 4(b), where small gaps are opened at $X$ and $Y$ and there is only the exact phase (with respect to non-Hermitian chiral symmetry) [42]. The band structure for $\gamma = 3.9$ is also shown in Figs. 4(c) and 4(d), where two EP2 loops shown earlier in Fig. 2(a) manifest clearly. We mention that similar evolution as shown in Fig. 4 could occur in one-dimensional systems such as the PT-symmetric model $H(k) = k\sigma_x + \alpha\sigma_y + i\sigma_z$, where $\sigma_i$ with $i = x, y, z$ are Pauli matrices. The bands are given by $E(k) = \pm\sqrt{k^2 + \alpha^2 - 1}$. The system possesses both PT-broken and PT-exact phases when $\alpha < 1$, whereas it contains only PT-exact phase with a gap when $\alpha > 1$.

### III. Systems with NNN hoppings
### A. Discrete EP2s

Now we consider the system with NNN hoppings. The NNN hopping $w$ couples the "A" and "C" sites as marked in Fig. 1(a), and enters the Hamiltonian via $T_{AC} = 4w\cos\left(\frac{k_x}{2}\right)\cos\left(\frac{k_y}{2}\right)$ in Eq. (1). All the $C_{4v}T$ group symmetries hold for $w \neq 0$. The triple degeneracy at $M$ remains intact for the Hermitian NNN system because $T_{AC} = 0$. The non-Hermitian chiral symmetry is no longer preserved for the non-Hermitian NNN system, which is indicated by violation the constraints given by Eqs. (2) and (3).

Instead of an EP loop, only discrete EP2s appear in the BZ when non-Hermiticity ($\gamma \neq 0$) is introduced. For illustration, the band structure is shown for $\gamma = 1$ and $w = 0.2$ in Figs. 5(a) and 5(b), where only a quarter of BZ is shown utilizing the inversion symmetry and mirror symmetry, namely Eq. (S4) [42]. There are only discrete EP2s, which are marked by 3 dots, black, cyan, and magenta, in Figs. 5(a) and 5(b). The yellow lines in Fig. 5(c) represent where $Re[E_i(\vec{k})] = Re[E_{j \neq i}(\vec{k})]$ and the green lines represent where $Im[E_i(\vec{k})] = Im[E_{j \neq i}(\vec{k})]$. And the thick (thin) lines mean that three (two) eigenvalues have identical real or imaginary parts. The three meeting points between yellow and green lines correspond to the 3 EP2s in Figs. 5(a) and 5(b). Thus there are 8 EP2s in the entire BZ as marked by dots in Fig. 5(d).

We now use the discriminant $\Delta(\vec{k})$ function to locate the EP2s. With $w \neq 0$, the characteristic polynomial is $f(E, \vec{k}) = E^3 + a_1 E + a_0$, where

$$a_1 = \gamma^2 - (T_{AB}^2 + T_{BC}^2 + T_{AC}^2),$$
$$a_0 = i\gamma(T_{BC}^2 - T_{AB}^2) - 2T_{AB}T_{BC}T_{AC}, \quad (12)$$

which reduces to Eq. (5) when $w = 0$. The discriminant has the following form,

$$\Delta(\vec{k}) = g_6 \gamma^6 + g_4 \gamma^4 + g_2 \gamma^2 + ig_1 \gamma + g_0, \quad (13)$$

where the coefficients are

$$g_6 = -4,$$
$$g_4 = 12(T_{AB}^2 + T_{AC}^2 + T_{BC}^2),$$
$$g_2 = 3(T_{AB}^2 - 2T_{AC}^2 - 5T_{BC}^2)(5T_{AB}^2 + 2T_{AC}^2 - T_{BC}^2),$$
$$g_1 = 108 T_{AB} T_{AC} T_{BC} (T_{BC}^2 - T_{AB}^2),$$
$$g_0 = 4(T_{AB}^2 + T_{AC}^2 + T_{BC}^2)^3 - 108 T_{AB}^2 T_{AC}^2 T_{BC}^2.$$

$$(14)$$

The discriminant $\Delta(\vec{k})$ in Eq. (13) is a complex function of $\vec{k}$ due to the breaking of the non-Hermitian chiral symmetry, which reduces to Eq. (6) when $w = 0$. We can write it as $\Delta = \Delta_R + i\Delta_I$, where both

$$\Delta_R = g_6 \gamma^6 + g_4 \gamma^4 + g_2 \gamma^2 + g_0 \quad (15)$$

and

$$\Delta_I = g_1 \gamma \quad (16)$$

are real functions of $\vec{k}$ in the BZ.

Just like the EP3s in Sec. II, now two constraints, i.e., $\Delta_R = 0$ and $\Delta_I = 0$, need to be satisfied simultaneously so that an EP can occur. Consequently only discrete EP2s are possible. For illustration, the solutions to $\Delta_R = 0$ and $\Delta_I = 0$ are plotted in Fig. 5(d) as navy loop and green lines, respectively. Their intersections are EP2s, denoted by colored dots. The discrete EP2s for the NNN system are restricted to high-symmetry lines $\Gamma M, XM$, and $YM$. In particular, the four EP2s located on $XM$ and $YM$ are inherited from the NN system, because $T_{AC} = 4w \cos\left(\frac{k_x}{2}\right) \cos\left(\frac{k_y}{2}\right)$ vanishes on $XM$ and $YM$. When $\gamma$ is increased from 0, the EP2s emanate from $M$ and move along the high-symmetry lines until they get annihilated in fours at $\Gamma$ or in pairs at $X$ and $Y$. EP2s may not exist when $\Delta_R = 0$ and $\Delta_I = 0$ are incompatible for certain system parameters.

For EP3s to exist in the NNN systems, three constraints, i.e., $\Delta_R = 0$, $\Delta_I = 0$ and $a_1 = 0$, need to be satisfied. That is generally impossible because there are three independent equations but only two variables, i.e., $k_x, k_y$. This is reflected by the absence of common intersections between contour lines of $\Delta_R = 0$, $\Delta_I = 0$ and $a_1 = 0$ [42].

### B. Topological protection and evolution of EPs

The complex nature of $\Delta(\vec{k}) = \Delta_R + i\Delta_I$ prompts us to define a vector field

$$\vec{D}(\vec{k}) = (\Delta_R, \Delta_I). \tag{17}$$

The EP2s appear at the zeros of the vector field $\vec{D}(\vec{k})$. A generalized winding number, referred to as a discriminant number, can be defined for each discrete EP2 [27],

$$\nu(\vec{k}^{EP}) = \frac{i}{2\pi} \oint_{C(\vec{k}^{EP})} d\vec{k} \cdot \nabla_{\vec{k}} \ln \Delta(\vec{k}), \tag{18}$$

where $C(\vec{k}^{EP})$ denotes a circular loop encircling an EP at $\vec{k}^{EP}$. The discriminant number is equal to the vorticity defined in Ref. [21] summed over all pairs of distinct bands [27]. For illustration, the vector field $\vec{D}(\vec{k})$ is plotted in Fig. 6(a) for $\gamma = 1, w = 0.2$ and Fig. 6(b) for $\gamma = 1, w = 0.295$, which contain 8 and 12 EP2s, respectively, as marked by blue dots. The vector field $\vec{D}(\vec{k})$ surrounding each EP2 is highlighted by red arrows.

The EP2s on the $\Gamma M$ segments are found to be either a vortex or a saddle point [51], corresponding to a discriminant number of $-1$ or $1$, which is confirmed by numerical calculations. Green and cyan circles mark the EP2s with $\nu = 1$ and $\nu = -1$, respectively. There is a sign difference between discriminant numbers of two EP2s that are mirror-symmetric about $k_x = \pi$ or $k_y = \pi$. This is due to $\Delta(2\pi - k_x, k_y) = \Delta(k_x, 2\pi - k_y) = \Delta(k_x, k_y)$ inferred from Eqs. (13) and (14). The same relation immediately leads to $\nu(\vec{k}^{EP}) = 0$ for the EP2s on $XM$ and $YM$ segments. We note that symmetries are assumed to be absent in Ref. [27], which is different from the situation here.

The discriminant number defined in Eq. (18) can be recast as $\nu(\vec{k}^{EP}) = \frac{i}{2\pi} \oint_{C(\vec{k}_{EP})} \frac{d \ln \Delta}{d\theta} d\theta$, which counts the winding number of $Arg(\Delta)$ around $\vec{k}_{EP}$, where $\theta$ is the polar angle on the loop $C$. In the following, we focus on two EP2s, one at $\vec{k}_1^{EP} \approx (0.7, 0.7)\pi$ with $\nu(\vec{k}_1^{EP}) = 1$ and the other at $\vec{k}_2^{EP} = \left(2 \cos^{-1}\left(-\frac{4}{\gamma}\right), \pi\right) \approx (1.2, 1)\pi$ with $\nu(\vec{k}_2^{EP}) = 0$. The variation of $Arg(\Delta)$ with increasing $\theta$ is shown in Fig. 6(c), which explicitly demonstrates $\nu(\vec{k}_1^{EP}) = 1$. Similarly $\nu(\vec{k}_2^{EP}) = 0$ is demonstrated in Fig. 6(d). The steep jumps near $\theta = \pi/2, 3\pi/2$ result from the facts that $|Im(\Delta)|$ is small and $Re(\Delta)$ changes sign there. The mirror symmetry is also reflected in Fig. 6(d) where the curve is symmetric about $\theta = \pi$.

Nonzero discriminant numbers of EP2s, which can be viewed as quantized topological charges, guarantee their stability against perturbations. The fermion doubling theorem for EPs dictates the neutrality of the charges so that the sum of the discriminant numbers for all the EPs in the BZ must vanish [27], which is demonstrated in Figs. 6(a) and 6(b). Moreover, the neutrality should be preserved when EPs are created or annihilated by tuning system parameters, which is exemplified by the evolution from Fig. 6(a) to Fig. 6(b) during which two EP2s with $\nu = 1$ and two EP2s with $\nu = -1$ are created from $\Gamma$.

To be more specific, we tune $w$ from 0.2 to 0.302 continuously and show the creation and annihilation behaviors of the discrete EP2s in Fig. 7. The locations of EP2s for 4 typical values of $w$ are shown in Fig. 7(a), where 4 EP2s with $\nu = 0$ remain fixed on $k_x = \pi$ and $k_y = \pi$ due to their independence from $w$. And the EP2s with $\nu \neq 0$ evolve on $\Gamma M$ lines, and their number changes from 4 to 8 and then to 0 due to creation and annihilation. Four additional EP2s are created from $\Gamma$ at $w \approx 0.29$, as is shown in Fig. 7(a). Further increasing $w$ causes the two EP2s on each $\Gamma M$ segments to approach each other and finally annihilate in pairs when $w$ is slightly over 0.3.

The band structures along $\Gamma M$ for $w = 0.25, 0.29, 0.925, 0.3, 0.302$ are shown in rows in Figs. 7(b) and 7(c), where only the two bands associated with the EP2s are plotted. The first four rows are plotted with colors taken in accordance with Fig. 7(a). There is a gap opening in the imaginary part of the bands for $w = 0.302$, which indicates the annihilation of two EP2s. The evolution of the EP2s with $\nu \neq 0$ is explicitly demonstrated in the band structures.

The EP2s with $\nu \neq 0$ are robust against changing $w$, that is, an EP2 remains intact during the process until it gets annihilated by another EP2 with an opposite charge, as is demonstrated in Fig. 7. Such topological protection originates from the fact that the discriminant number is an integer and has to remain unchanged, because they should vary continuously when the system is changed smoothly. The topologically protected robustness of EP2s with $\nu \neq 0$ against other perturbations will be discussed in Sec. IV.

### C. Anisotropic behavior at discrete EP2s

We have shown in Fig.3 that the EPs residing on the EP loop exhibit anisotropic behaviors. We found that similar anisotropic behaviors also occur to the discrete EP2s in Fig. 5(d). This is very surprising because there is no longer a direction which is obviously as special as the tangent direction at an EP loop in Fig. 3(a). This anisotropy is due to the particular band structure shown in Fig. 5 which inherited features of the band structure of the NN system shown in Figs. 1(c) and 1(d). Thus discrete EP2s could exhibit both anisotropic and isotropic behaviors recalling the Dirac-like cones shown in Fig. 4(a).

### IV. Effects of perturbation

We have shown the interesting EP behaviors in both the NN system and the NNN system. Now we discuss the effect of perturbations on the EPs.

For the NN system, the presence of EP2 arcs and EP3s necessitates the non-Hermitian chiral symmetry. Now we introduce a chiral-symmetry-breaking perturbation so that the Hamiltonian becomes $H' = H + \Delta H$, where $\Delta H = \delta s_z = \delta\, Diag(1,0,-1)$. The EP loop for the unperturbed NN system with $\gamma = 2$ is shown as the black curve in Fig. 8(a). An infinitesimal perturbation wipes out all the EP2 arcs immediately and splits each EP3 into two EP2s, due to the breaking of the non-Hermitian chiral symmetry. Continuously increasing $\delta$ traces out the moving trajectories of the 8 EP2s. The EP2s for $\delta = 0.15, 0.3, 0.5, 0.7$ are shown by circles with different colors. An EP2 on a red (blue) trajectory has $\nu = 1$ ($\nu = -1$). The total charge remains neutral during the evolution as dictated by the fermion doubling theorem [27]. Pairs of EP2s with $\nu = \pm 1$ will meet and annihilate at the BZ boundaries when $\delta$ is increased to some critical value. Thus the discrete EP2s with $\nu = \pm 1$ are robust against chiral-symmetry-breaking perturbations though the EP loop is fragile.

For the NNN system with $\gamma = 1$ and $w = 0.2$, the 8 discrete EP2s are shown as black dots in Fig. 8(b). We still introduce a perturbation $\Delta H = \delta s_z$. The 8 EP2s exhibit three types of behaviors in the presence of a small $\delta$: (1) The four EP2s located at the BZ diagonals with $\nu = \pm 1$ shift slightly due to their topological robustness; (2) the two EP2s located at $k_y = \pi$ with $\nu = 0$ are wiped out; and (3) the two EP2s located at $k_x = \pi$ with $\nu = 0$ split into two EP2s. The results (2) and (3) mean that the four EP2s with $\nu = 0$ are fragile under the perturbation that breaks the non-Hermitian chiral symmetry and are either wiped out or split into EP2s with $\nu = \pm 1$, indicating the absence of topological protection. We mention that the EP2s with $\nu = 0$ are robust against the non-Hermitian chiral-symmetric perturbations because they originate from the preserved the non-Hermitian chiral symmetry along $XM$ and $YM$. Increasing $\delta$ continuously traces out the moving trajectories of the EP2s as shown in Fig. 8(b), where the red and blue colors represent $\nu = 1$ and $\nu = -1$, respectively. We emphasize that the EP2s with $\nu = \pm 1$ which result from the spitting are robust against further increase in $\delta$. Thus, the EP2s with $\nu = 0$ make a stark contrast to those EP2s with $\nu = \pm 1$ when responding to perturbations that breaks the non-Hermitian chiral symmetry. All the EP2s get annihilated in pairs at the BZ boundary for some large $\delta$. The neutrality of topological charges is always preserved. We mention that introducing further-neighbor hoppings does not bring qualitative difference compared to the NNN system.

To conclude, discrete EP2s are ubiquitous in the non-Hermitian Lieb lattice system because there are as many degrees of freedom as the constraints, whereas the EP3s and EP loop are fragile against breaking of the non-Hermitian chiral symmetry. The robustness of discrete EP2s against perturbations is well characterized by a nonzero discriminant number.

### V. Realization by coupled waveguides

We propose that the non-Hermitian Lieb lattice could be realized by evanescently coupled optical waveguides with loss/gain added to "A"/"C" waveguides as shown in Fig. 9(a). The evolution of optical field in the $z$ direction can be well described by a discrete Schrödinger equation [39,40],

$$-i\frac{d}{dz}\varphi_{\vec{m}} = \beta_{\vec{m}}\varphi_{\vec{m}} + \sum_{\vec{n}} V_{\vec{m},\vec{n}}\varphi_{\vec{n}}, \qquad (19)$$

where $\varphi_{\vec{m}}$ denotes the field amplitude at the waveguide indexed by $\vec{m}$, which specifies the location of the waveguide in the $xy$-plane. All waveguides are assumed to be identical and have the same propagation constant $\beta_0$ without loss/gain. The propagation constants for "A" and "C" waveguides become $\beta_0 + i\gamma$ and $\beta_0 - i\gamma$ in the presence of loss/gain. Assuming $\varphi_{\vec{m}}(z) = \phi_{\vec{m}} e^{i\beta z}$, Eq. (19) becomes

$$\beta\phi_{\vec{m}} = \beta_{\vec{m}}\phi_{\vec{m}} + \sum_{\vec{n}} V_{\vec{m},\vec{n}}\phi_{\vec{n}}, \qquad (20)$$

which mimics a tight-binding model with the propagation constant $\beta$ substituted for the eigen-energy $E$ in $H|\psi\rangle = E|\psi\rangle$. For the Lieb photonic lattice, Eq. (20) can be recast into $M(\vec{k})\Phi =$

$\beta\Phi$, where $\Phi = (u_A, u_B, u_C)^T$ denotes the cell-periodic Bloch function and $M(\vec{k})$ takes the same form as $H(\vec{k})$ in Eq. (1) except for an identity matrix $\beta_0 \mathbb{I}$.

For illustration, we take the lattice constant and the cylinder radius to be $44\mu m$ and $10\mu m$, respectively. The working wavelength is $\lambda = 1\mu m$. The refractive indices for the background medium and the waveguides are $n_1 = 1.473$ and $n_2 = 1.4737$. The refractive indices for the "A" and "C" waveguides are respectively $n_2 + i\eta$ and $n_2 - i\eta$ where $\eta = 10^{-5}$, which correspond to loss and gain (note the sign difference from the tight-binding model). The band structure of $\beta$ is shown in Figs. 9(b) and 9(c), where we see an excellent agreement between the dots from the full-wave simulation and the red curves from fitting to the NNN tight-binding model. The discrete EP2s appear at $\Gamma M$, MX, and MY, which is reminiscent of Fig. 6(a) where there are 8 EP2s in the BZ. The couplings from further neighbors are negligibly weak and their presence should only shift the EP2s with $\nu \neq 0$ slightly due to the topologically protected robustness. Similarly the more interesting evolution behavior of EP2s shown in Fig. 7 can also be realized on the coupled waveguides platform. Even the EP2 loop and EP3s could be traced out by specially tailoring system parameters so that the NNN hoppings are negligible compared to the NN hoppings.

The material loss of the fabricated waveguide array is often negligible. Practically gain/loss could be replaced by differential loss added to different waveguides. The loss could be engineered, for example, by creating breaks in waveguides [38]. In addition to the coupled waveguides, coupled acoustic resonators [12,13,52] are also good candidates to realize the non-Hermitian Lieb lattice.

## VI. Conclusions

We showed the interesting phenomenon that EP3 cusps join EP2 arcs and form an EP loop in the non-Hermitian Lieb lattice with NN hoppings. The existence of the EP loop and EP3s results from the realness of the discriminant guaranteed by the non-Hermitian chiral symmetry. The EP loop expands with the increasing non-Hermiticity until the EPs vanish at a critical value of non-Hermiticity. The remarkable anisotropic behaviors for both EP2s and EP3s are also discussed. Another interesting phenomenon is the occurrence of Dirac-like cones with EPs residing at the cone vertices. The EP loop is found to be fragile against perturbations that break the non-Hermiticity chiral symmetry. In the presence of NNN hoppings, the EP loop is wiped out and only several discrete EP2s survive. The EP2s with nonzero discriminant number (i.e., $\nu \neq 0$) is shown to possess topologically protected robustness, unless two EP2s with opposite discriminant numbers meet and annihilate each other. An EP2 with $\nu = 0$ may either be removed or get split into two EP2s with $v = \pm 1$ by perturbations. We found that the discrete EP2s also exhibit anisotropic behaviors. The experimental realization of the non-Hermitian Lieb lattice by coupled waveguides is discussed and demonstrated by the simulation results.

## ACKNOWLEDGMENTS

This work is supported by Hong Kong RGC (16303119, AoE/P-02/12, N_HKUST608/17). It is also supported by the National Natural Science Foundation of China (Grant No. 11847205) and China Postdoctoral Science Foundation (Grant No. 2018M630597). K.D. acknowledges funding from the Gordon and Betty Moore Foundation.

[1]  L. Feng, R. El-Ganainy, and L. Ge, *Non-Hermitian Photonics Based on Parity–Time Symmetry*, Nature Photonics **11**, 752 (2017).
[2]  R. El-Ganainy, K. G. Makris, M. Khajavikhan, Z. H. Musslimani, S. Rotter, and D. N. Christodoulides, *Non-Hermitian Physics and PT Symmetry*, Nature Physics **14**, 11 (2018).
[3]  Ş. K. Özdemir, S. Rotter, F. Nori, and L. Yang, *Parity–Time Symmetry and Exceptional Points in Photonics*, Nature Materials **18**, 783 (2019).
[4]  M.-A. Miri and A. Alù, *Exceptional Points in Optics and Photonics*, Science **363**, eaar7709 (2019).
[5]  C. Dembowski, B. Dietz, H.-D. Gräf, H. L. Harney, A. Heine, W. D. Heiss, and A. Richter, *Observation of a Chiral State in a Microwave Cavity*, Phys. Rev. Lett. **90**, 034101 (2003).
[6]  M. V. Berry, *Physics of Nonhermitian Degeneracies*, Czech. J. Phys. **54**, 1039 (2004).
[7]  J. Doppler, A. A. Mailybaev, J. Böhm, U. Kuhl, A. Girschik, F. Libisch, T. J. Milburn, P. Rabl, N. Moiseyev, and S. Rotter, *Dynamically Encircling an Exceptional Point for Asymmetric Mode Switching*, Nature (London) **537**, 76 (2016).
[8]  X.-L. Zhang, S. Wang, B. Hou, and C. T. Chan, *Dynamically Encircling Exceptional Points: In Situ Control of Encircling Loops and the Role of the Starting Point*, Phys. Rev. X **8**, 021066 (2018).
[9]  H. Hodaei, A. U. Hassan, S. Wittek, H. Garcia-Gracia, R. El-Ganainy, D. N. Christodoulides, and M. Khajavikhan, *Enhanced Sensitivity at Higher-Order Exceptional Points*, Nature (London) **548**, 187 (2017).
[10] K. Ding, Z. Q. Zhang, and C. T. Chan, *Coalescence of Exceptional Points and Phase Diagrams for One-Dimensional PT-Symmetric Photonic Crystals*, Phys. Rev. B **92**, 235310 (2015).
[11] B. Zhen, C. W. Hsu, Y. Igarashi, L. Lu, I. Kaminer, A. Pick, S.-L. Chua, J. D. Joannopoulos, and M. Soljačić, *Spawning Rings of Exceptional Points out of Dirac Cones*, Nature (London) **525**, 354 (2015).

[12] K. Ding, G. Ma, M. Xiao, Z. Q. Zhang, and C. T. Chan, *Emergence, Coalescence, and Topological Properties of Multiple Exceptional Points and Their Experimental Realization*, Phys. Rev. X **6**, 021007 (2016).

[13] K. Ding, G. Ma, Z. Q. Zhang, and C. T. Chan, *Experimental Demonstration of an Anisotropic Exceptional Point*, Phys. Rev. Lett. **121**, 085702 (2018).

[14] X.-X. Zhang and M. Franz, *Non-Hermitian Exceptional Landau Quantization in Electric Circuits*, Phys. Rev. Lett. **124**, 046401 (2020).

[15] T. Stehmann, W. D. Heiss, and F. G. Scholtz, *Observation of Exceptional Points in Electronic Circuits*, J. Phys. A: Math. Gen. **37**, 7813 (2004).

[16] B. Peng, Ş. K. Özdemir, M. Liertzer, W. Chen, J. Kramer, H. Yılmaz, J. Wiersig, S. Rotter, and L. Yang, *Chiral Modes and Directional Lasing at Exceptional Points*, PNAS **113**, 6845 (2016).

[17] L. Jin and Z. Song, *Incident Direction Independent Wave Propagation and Unidirectional Lasing*, Phys. Rev. Lett. **121**, 073901 (2018).

[18] J. Zhang, B. Peng, Ş. K. Özdemir, K. Pichler, D. O. Krimer, G. Zhao, F. Nori, Y. Liu, S. Rotter, and L. Yang, *A Phonon Laser Operating at an Exceptional Point*, Nat. Photonics **12**, 479 (2018).

[19] T. E. Lee, *Anomalous Edge State in a Non-Hermitian Lattice*, Phys. Rev. Lett. **116**, 133903 (2016).

[20] D. Leykam, K. Y. Bliokh, C. Huang, Y. D. Chong, and F. Nori, *Edge Modes, Degeneracies, and Topological Numbers in Non-Hermitian Systems*, Phys. Rev. Lett. **118**, 040401 (2017).

[21] H. Shen, B. Zhen, and L. Fu, *Topological Band Theory for Non-Hermitian Hamiltonians*, Phys. Rev. Lett. **120**, 146402 (2018).

[22] S. Yao and Z. Wang, *Edge States and Topological Invariants of Non-Hermitian Systems*, Phys. Rev. Lett. **121**, 086803 (2018).

[23] S. Yao, F. Song, and Z. Wang, *Non-Hermitian Chern Bands*, Phys. Rev. Lett. **121**, 136802 (2018).

[24] F. K. Kunst, E. Edvardsson, J. C. Budich, and E. J. Bergholtz, *Biorthogonal Bulk-Boundary Correspondence in Non-Hermitian Systems*, Phys. Rev. Lett. **121**, 026808 (2018).

[25] D. S. Borgnia, A. J. Kruchkov, and R.-J. Slager, *Non-Hermitian Boundary Modes and Topology*, Phys. Rev. Lett. **124**, 056802 (2020).

[26] Y.-X. Xiao, Z.-Q. Zhang, Z. H. Hang, and C. T. Chan, *Anisotropic Exceptional Points of Arbitrary Order*, Phys. Rev. B **99**, 241403 (2019).

[27] Z. Yang, A. P. Schnyder, J. Hu, and C.-K. Chiu, *Fermion Doubling Theorems in 2D Non-Hermitian Systems for Fermi Points and Exceptional Points*, ArXiv:1912.02788 (2020).

[28] K. Moors, A. A. Zyuzin, A. Yu. Zyuzin, R. P. Tiwari, and T. L. Schmidt, *Disorder-Driven Exceptional Lines and Fermi Ribbons in Tilted Nodal-Line Semimetals*, Phys. Rev. B **99**, 041116 (2019).

[29] Y. Xu, S.-T. Wang, and L.-M. Duan, *Weyl Exceptional Rings in a Three-Dimensional Dissipative Cold Atomic Gas*, Phys. Rev. Lett. **118**, 045701 (2017).

[30] T. Yoshida, R. Peters, N. Kawakami, and Y. Hatsugai, *Symmetry-Protected Exceptional Rings in Two-Dimensional Correlated Systems with Chiral Symmetry*, Phys. Rev. B **99**, 121101 (2019).

[31] R. Okugawa and T. Yokoyama, *Topological Exceptional Surfaces in Non-Hermitian Systems with Parity-Time and Parity-Particle-Hole Symmetries*, Phys. Rev. B **99**, 041202 (2019).

[32] J. C. Budich, J. Carlström, F. K. Kunst, and E. J. Bergholtz, *Symmetry-Protected Nodal Phases in Non-Hermitian Systems*, Phys. Rev. B **99**, 041406 (2019).

[33] X. Zhang, K. Ding, X. Zhou, J. Xu, and D. Jin, *Experimental Observation of an Exceptional Surface in Synthetic Dimensions with Magnon Polaritons*, Phys. Rev. Lett. **123**, 237202 (2019).

[34] K. Kimura, T. Yoshida, and N. Kawakami, *Chiral-Symmetry Protected Exceptional Torus in Correlated Nodal-Line Semimetals*, Phys. Rev. B **100**, 115124 (2019).

[35] H. Zhou, J. Y. Lee, S. Liu, and B. Zhen, *Exceptional Surfaces in PT-Symmetric Non-Hermitian Photonic Systems*, Optica, OPTICA **6**, 190 (2019).

[36] W. Tang, X. Jiang, K. Ding, Y.-X. Xiao, Z.-Q. Zhang, C. T. Chan, and G. Ma, *Exceptional Nexus with a Hybrid Topological Invariant*, Science **370**, 1077 (2020).

[37] A. Cerjan, M. Xiao, L. Yuan, and S. Fan, *Effects of Non-Hermitian Perturbations on Weyl Hamiltonians with Arbitrary Topological Charges*, Phys. Rev. B **97**, 075128 (2018).

[38] A. Cerjan, S. Huang, M. Wang, K. P. Chen, Y. Chong, and M. C. Rechtsman, *Experimental Realization of a Weyl Exceptional Ring*, Nat. Photonics **13**, 623 (2019).

[39] R. A. Vicencio, C. Cantillano, L. Morales-Inostroza, B. Real, C. Mejía-Cortés, S. Weimann, A. Szameit, and M. I. Molina, *Observation of Localized States in Lieb Photonic Lattices*, Phys. Rev. Lett. **114**, 245503 (2015).

[40] S. Mukherjee, A. Spracklen, D. Choudhury, N. Goldman, P. Öhberg, E. Andersson, and R. R. Thomson, *Observation of a Localized Flat-Band State in a Photonic Lieb Lattice*, Phys. Rev. Lett. **114**, 245504 (2015).

[41] S. M. Zhang and L. Jin, *Flat Band in Two-Dimensional Non-Hermitian Optical Lattices*, Phys. Rev. A **100**, 043808 (2019).

[42] See Supplemental Material at [URL will be inserted by publisher] for discussions of symmetries, derivations, supplemental figures, exact/broken phase separation by non-Hermitian chiral symmetry, etc., which includes *Refs [27, 45, 46, 53–58]*.

[43] M. Am-Shallem, R. Kosloff, and N. Moiseyev, *Exceptional Points for Parameter Estimation in Open Quantum Systems: Analysis of the Bloch Equations*, New J. Phys. **17**, 113036 (2015).

[44] D. Dizdarevic, D. Dast, D. Haag, J. Main, H. Cartarius, and G. Wunner, *Cusp Bifurcation in the Eigenvalue Spectrum of PT-Symmetric Bose-Einstein Condensates*, Phys. Rev. A **91**, 033636 (2015).

[45] K. Kawabata, K. Shiozaki, M. Ueda, and M. Sato, *Symmetry and Topology in Non-Hermitian Physics*, Phys. Rev. X **9**, 041015 (2019).

[46] Z. Oztas and C. Yuce, *Spontaneously Broken Particle-Hole Symmetry in Photonic Graphene with Gain and Loss*, Phys. Rev. A **98**, 042104 (2018).

[47] X. Cui, K. Ding, J.-W. Dong, and C. T. Chan, *Exceptional Points and Their Coalescence of PT-Symmetric Interface States in Photonic Crystals*, Phys. Rev. B **100**, 115412 (2019).

[48] X.-L. Zhang and C. T. Chan, *Hybrid Exceptional Point and Its Dynamical Encircling in a Two-State System*, Phys. Rev. A **98**, 033810 (2018).

[49] L. Jin, H. C. Wu, B.-B. Wei, and Z. Song, *Hybrid Exceptional Point Created from Type-III Dirac Point*, Phys. Rev. B **101**, 045130 (2020).

[50] I. Rotter, *A Non-Hermitian Hamilton Operator and the Physics of Open Quantum Systems*, J. Phys. A: Math. Theor. **42**, 153001 (2009).

[51] T. Needham, *Visual Complex Analysis* (Clarendon, Oxford, 2000).

[52] Y.-X. Xiao, G. Ma, Z.-Q. Zhang, and C. T. Chan, *Topological Subspace-Induced Bound State in the Continuum*, Phys. Rev. Lett. **118**, 166803 (2017).

[53] Daniel. B. Litvin, Magnetic Group Tables: 1-, 2- and 3-Dimensional Magnetic Subperiodic Groups and Magnetic Space Groups (International Union of Crystallography, Chester, England, Chester, England, 2013).

[54] K. Kawabata, T. Bessho, and M. Sato, Classification of Exceptional Points and Non-Hermitian Topological Semimetals, Phys. Rev. Lett. 123, 066405 (2019).

[55] J. W. Bruce and P. J. Giblin, Curves and Singularities: A Geometrical Introduction to Singularity Theory (Cambridge University Press, Cambridge, 1992).

[56] S. Bittner, B. Dietz, U. Günther, H. L. Harney, M. Miski-Oglu, A. Richter, and F. Schäfer, PT Symmetry and Spontaneous Symmetry Breaking in a Microwave Billiard, Phys. Rev. Lett. 108, 024101 (2012).

[57] K. Esaki, M. Sato, K. Hasebe, and M. Kohmoto, Edge States and Topological Phases in Non-Hermitian Systems, Physical Review B 84, 205128 (2011).

[58] S. Lieu, Topological Phases in the Non-Hermitian Su-Schrieffer-Heeger Model, Phys. Rev. B 97, 045106 (2018).

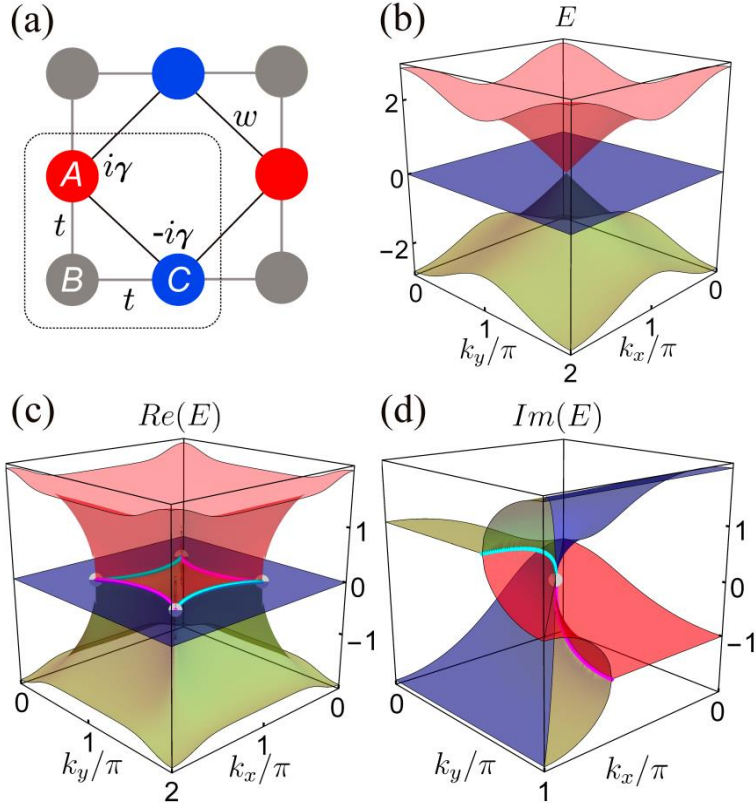

Fig. 1 (a) The non-Hermitian Lieb lattice with onsite gain/loss ($\pm i\gamma$). (b) The band structure of the Hermitian Lieb lattice ($\gamma = 0$) with NN hoppings exhibits a triple degeneracy at $M = (\pi, \pi)$. (c) The real part and (d) imaginary part of the band structure for the non-Hermitian Lieb lattice ($\gamma = 2$) with NN hoppings exhibit a star-shaped exceptional loop. For good visualization only a quarter of the Brillouin zone is shown for the imaginary part utilizing the symmetry $E(k_x, k_y) = E(2\pi - k_x, k_y) = E(k_x, 2\pi - k_y)$. The magenta and cyan arcs correspond to exceptional points of order 2, and the four cusps correspond to exceptional points of order 3.

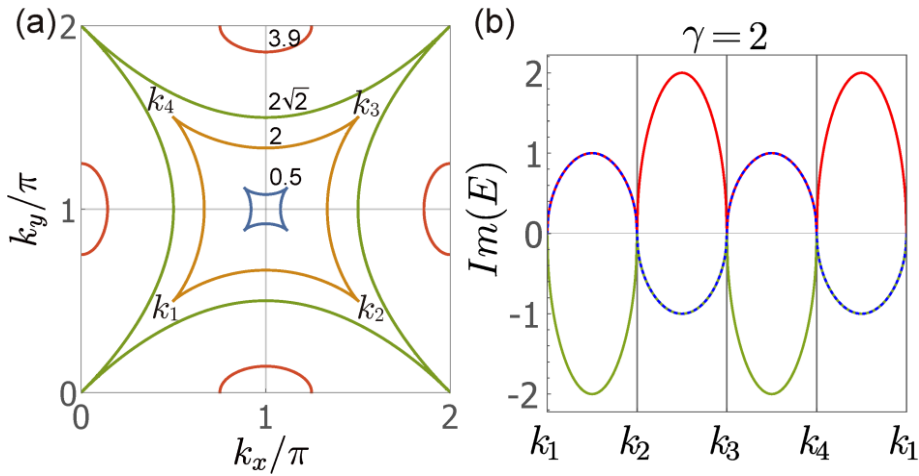

Fig. 2 (a) An exceptional loop emanates at $M$ for a small $\gamma$ and then expands with increasing $\gamma$. The values of $\gamma$ are labeled at each EP loop. (b) The bands along the EP loop for $\gamma = 2$ is shown, where the ticks $k_1, k_2, k_3, k_4$ denote the four EP3 cusps labeled in (a).

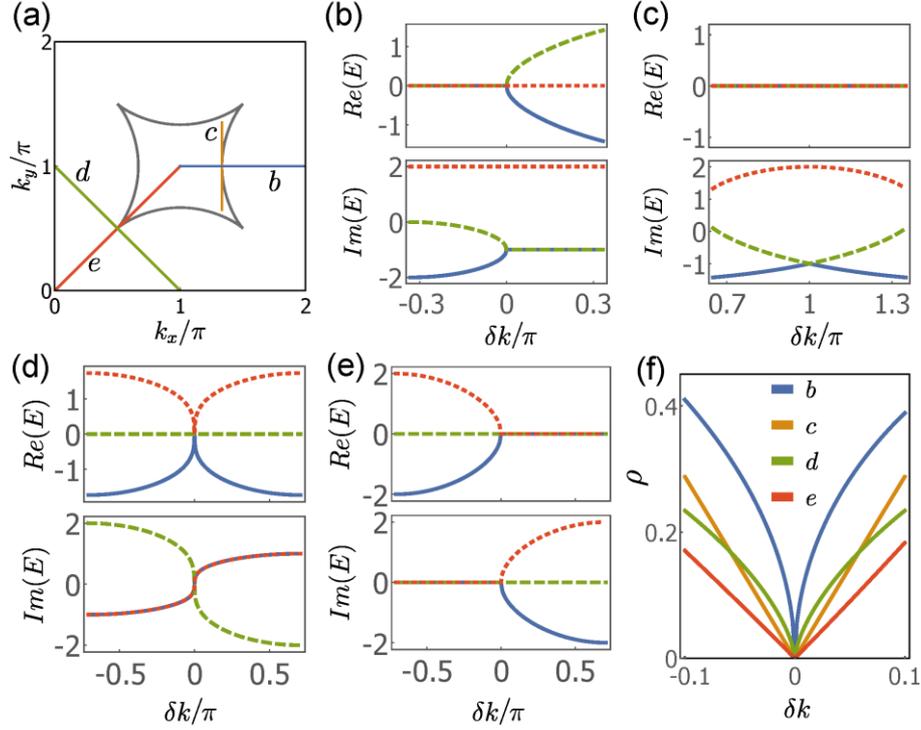

Fig. 3 The anisotropic behaviors at the EP2 at $\left(\frac{4}{3}, 1\right)\pi$ and the EP3 at $\left(\frac{1}{2}, \frac{1}{2}\right)\pi$. (a) The normal/tangent paths traversing the EP2 and EP3 are denoted by navy/orange and green/red lines. The labels "b", "c", "d" and "e" associate the paths with the subfigures (b)-(e). (b)-(e) The energy bands along the four paths in (a), where $\delta k$ denotes the small deviations from the EP2 or EP3. (f) The phase rigidity $\rho$ near the EP2/EP3 for the four paths has critical exponents $\frac{1}{2}, 1, \frac{2}{3}, 1$, respectively. The values of $\rho$ for the "c" and "e" paths are scaled up by a factor of 5.

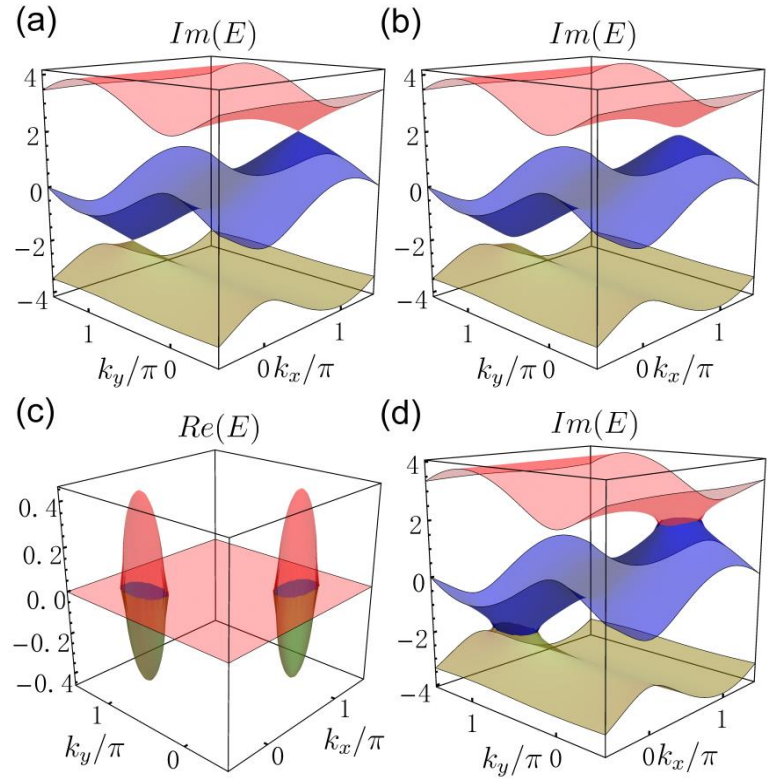

Fig. 4 (a) Two Dirac-like cones reside at $X$ and $Y$ in the purely imaginary bands when $\gamma = 4$. EP2s appear at the cone vertices. (b) The bands become gapped at $X$ and $Y$ when $\gamma = 4.01$. (c) The real part and (d) imaginary part of the bands exhibit two EP2 loops centered $X$ and $Y$ when $\gamma = 3.9$.

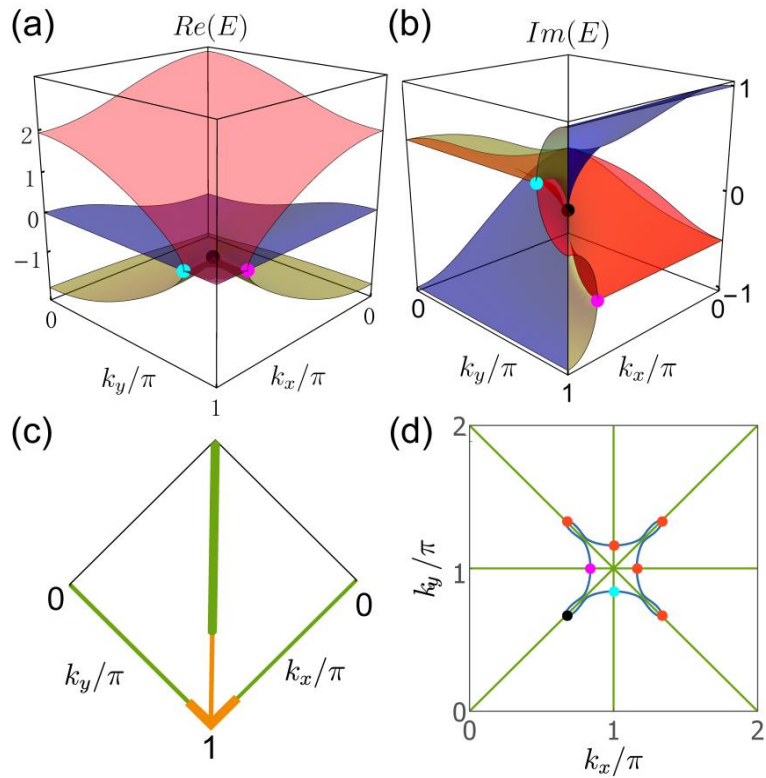

Fig. 5 The band structure and exceptional points in the Lieb lattice with NNN hoppings. (a) The real part and (b) imaginary part of the band structure with $\gamma = 1, w = 0.2$ are shown in a quarter of BZ. The three colored dots correspond to discrete EP2s. (c) The thick (thin) yellow lines represent where 3 (2) eigenvalues have the same real part. The green lines represent where eigenvalues have identical imaginary parts. Discrete EP2s occur at the three joints. (d) The contour lines of $\text{Re}(\Delta) = 0$ and $\text{Im}(\Delta) = 0$ in the BZ are represented by a navy curved loop and green lines, respectively. The 8 intersections marked by dots are EP2s. The black, cyan, magenta dots correspond to the three dots in (a) and (b).

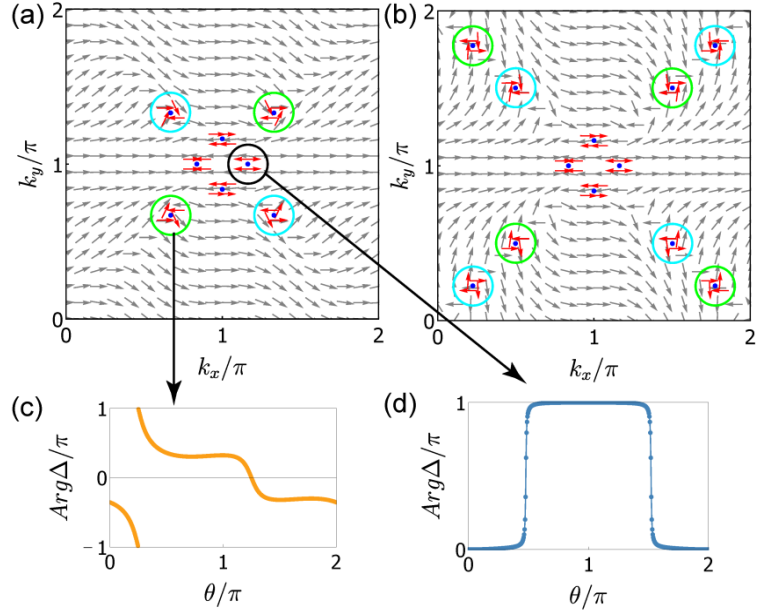

Fig. 6 The vector field distributions $\vec{D} = (\Delta_R, \Delta_I)$ are shown for two typical cases: (a) $w = 0.2$ and (b) $w = 0.295$. $\gamma = 1$ is assumed. The field distributions near the EP2s are highlighted by red arrows. The green and cyan circles mark the EP2s with $\nu = 1$ and $\nu = -1$, respectively. The four EP2s at $k_x = \pi$ and $k_y = \pi$ have a discriminant number of zero. The variations of $\text{Arg}(\Delta)$ around (c) an EP2 with $\nu = 1$ and (d) an EP2 with $\nu = 0$ are shown.

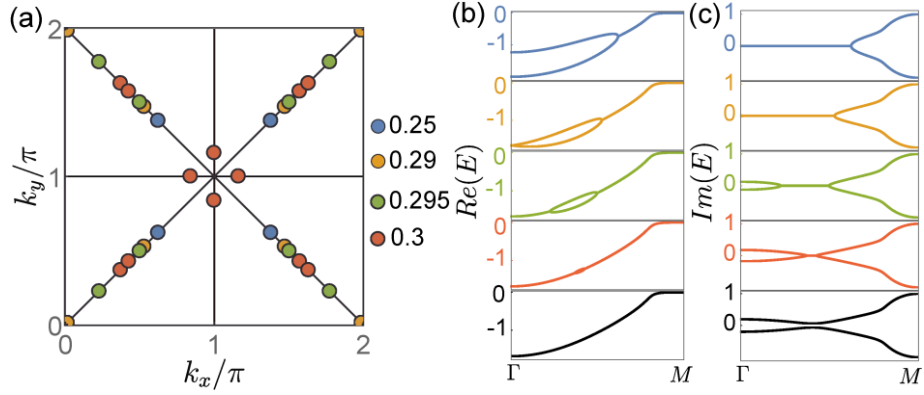

Fig. 7 The evolution of the discrete EP2s with increasing NNN hopping $w$. (a) The discrete EP2s are shown for 4 typical values of NNN hoppings: $w = 0.25, 0.29, 0.295, 0.3$. The 4 EP2s at $k_x = \pi$ and $k_y = \pi$ remain invariant when $w$ is varied. There are 4, 8, 8, 8 EP2s at the BZ diagonals for the 4 cases, respectively. The EP2s are created in fours at $\Gamma$ (when $w \approx 0.29$) or annihilated in pairs on $\Gamma M$ (when $w$ is slightly above 0.3) during the increasing of $w$. (b) The real parts and (c) imaginary parts of the bands along $\Gamma M$ are shown for $w = 0.25, 0.29, 0.295, 0.3$ and $0.302$ in five rows. Only the two bands associated with the EP2s are plotted.

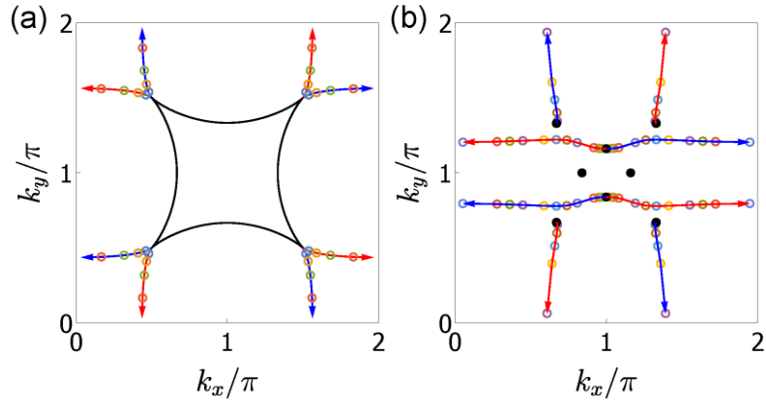

Fig. 8 The evolution of the EP loop (NN system) and the discrete EP2s (NNN system) under the perturbation $\Delta H = \delta\, Diag(1,0,-1)$. (a) The EP loop for the unperturbed NN system with $\gamma = 2$ is shown as the black curve. The perturbation wipes out all EP2s and split each EP3 into two EP2s. The increase of $\delta$ traces out the trajectories for the 8 split EP2s. The discriminant number is $\nu = 1\ (-1)$ for an EP2 on a red (blue) trajectory. The colored circles mark different values of $\delta$. (b) The EP2s for the unperturbed NNN system with $\gamma = 1, w = 0.2$ are denoted by the 8 black dots. When perturbed,

the 4 EP2s with $v = \pm 1$ (at the diagonals) move towards $k_y = 0$ and $k_y = 2\pi$. The two $v = 0$ EP2s at $k_y = \pi$ disappear immediately when perturbed, whereas the two $v = 0$ EP2s at $k_x = \pi$ split into two EP2s with nonzero discriminant number ($v = \pm 1$). The EP2s finally annihilate in pairs at the BZ boundaries when $\delta$ is increased to certain values.

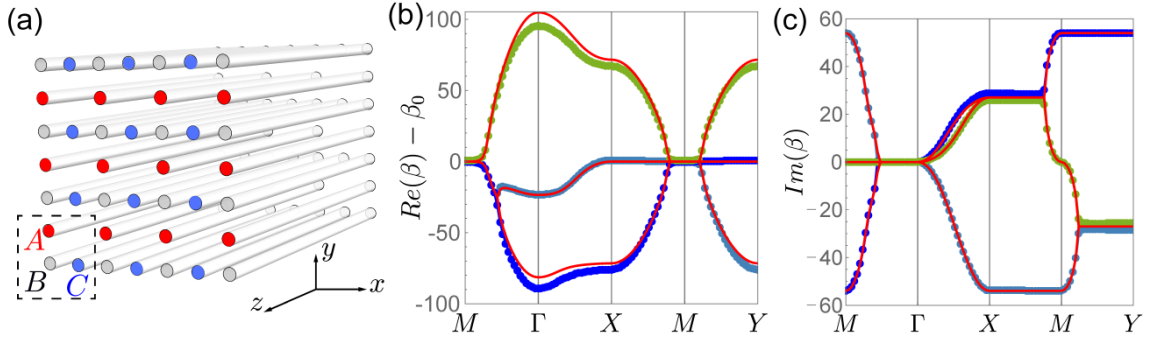

Fig. 9 (a) A coupled optical waveguide array forms a non-Hermitian Lieb lattice, where loss/gain is denoted by red/blue colors. (b) The real part and (c) imaginary part of the band structure of the mode propagation constant $\beta$ as a function of the quasi-momentum $\vec{k}$ are shown. The dots are from the full-wave simulation with the lattice constant and cylinder radius set to be $44\mu m$ and $10\mu m$, respectively. The refractive indices are $n_1, n_2 + i\eta, n_2, n_2 - i\eta$ for the background medium, "A" cylinders, "B" cylinders and "C" cylinders, respectively, where $n_1 = 1.473, n_2 = 1.4737$ and $\eta = 10^{-5}$. The red curves are the band structure of the fitted tight-binding model with NNN hoppings.